\documentclass{PoS}
  
\title{Energy dependence of $\phi$-meson production and elliptic flow 
in Au+Au collisions at STAR}

\ShortTitle{Energy dependence of $\phi$-meson production and elliptic flow 
in Au+Au collisions at STAR}

\author{\speaker{Md. Nasim (for the STAR collaboration)}\\
  School of Physical Science\\
        National Institute of Science Education and Research\\
Bhubaneswar-751 005, India\\
        E-mail: \email{nasim@rcf.rhic.bnl.gov}}

\abstract{We present, the measurements of the $\phi$-meson production and elliptic flow ($v_{2}$) at mid-rapidity in Au + Au
  collisions at $\sqrt{s_{NN}}$ = 7.7 - 200 GeV 
  using the STAR detector in the years 2010 and 2011. The energy dependence of nuclear
  modification factors ($R_{\rm{CP}}$) is presented. At the intermediate
  transverse momentum ($p_{T}$) the $\Omega/\phi$ ratio shows a different
  trend for $\sqrt{s_{NN}}$ = 11.5 GeV  compared to higher beam energies.
This may suggest
  change of particle production mechanism at $\sqrt{s_{NN}}$ = 11.5
  GeV. 
The number-of-constituent quark (NCQ) scaling of $v_{2}$ has been studied at various
beam energies. The NCQ scaling holds for all identified particles for
$\sqrt{s_{NN}}$ $\geq$ 19.6 GeV, which can be considered as an evidence of
partonic collectivity. We observe 
at $\sqrt{s_{NN}}$ = 7.7 and 11.5 GeV,  the $\phi$-meson $v_{2}$ falls off the
trend from the other hadrons at highest measured $p_{ T}$ 
values by 1.8$\sigma$ and 2.3$\sigma$, respectively. This may
indicate that the hadronic interaction plays an important role at lower beam energies.
  }

\FullConference{8th International Workshop on
Critical Point and Onset of Deconfinement,\\
		March 11 to 15, 2013\\
		Napa, California, USA}

\begin{document}

\section{Introduction}
The $\phi$ vector meson is the lightest bound state of $s$ and
$\bar{s}$ quarks. The interaction cross-section
of the $\phi$ meson  with
non-strange hadrons is expected to have a small value~\cite{white} and
therefore  its production should be less affected by the later
stage hadronic interactions in the evolution of the system formed in heavy-ion collisions. The
$\phi$ meson seems to freeze out early compared to other light hadrons 
($\pi$, $K$ and $p$)~\cite{white}. The life time of the $\phi$ meson is $\sim$ 42
fm/$c$. Because of longer life time the $\phi$ meson will mostly decay outside
the fireball and therefore its daughters will not have much time to
re-scatter in the hadronic phase.
The elliptic flow ($v_{2}$), a measure of the anisotropy in momentum
space, for $\phi$ meson can be used to probe the dynamics of the early
stage of heavy-ion collisions~\cite{hydro}. For the $\phi$-meson $v_{2}$, effect of later stage
hadronic interaction is small~\cite{BN,NBN}. Therefore, the $\phi$ meson can be considered as a clean
probe to study the QCD phase diagram in the Beam Energy Scan (BES)
program at the Relativistic Heavy Ion Collider (RHIC)~\cite{starnote}.

\section{Data sets and methods}
The results presented here are based on data collected at
$\sqrt{s_{NN}}$= 7.7, 11.5, 19.6, 27, 39,  62.4 and 200 GeV in Au+Au collisions by the
STAR detector for a minimum bias trigger in the years of
2010 and 2011.
The Time
Projection Chamber (TPC)
and Time of Flight (TOF) detectors 
with full $2\pi$ coverage were used for particle identification in the
central pseudo-rapidity ($\eta$) region ($|\eta|<$ 1.0). 
$\phi$ mesons are identified using the invariant mass technique from
their decay  to $K^{+} + K^{-}$ (branching ratio is 49.04 $\pm$ 0.6
$\%$). Mixed event technique has been used for combinatorial
background estimation~\cite{phi_plb_star}. The $\eta$-sub event plane
method~\cite{method} using TPC tracks
has been applied to measure the elliptic flow.
In this method, one defines the event flow vector for each
particle based on particles measured in the opposite hemisphere in
pseudo-rapidity ($\eta$).  
An $\eta$ gap of $|\eta| <$ 0.05 between positive and negative
pseudo-rapidity sub-events has been introduced to suppress non-flow
effects.

\section{Results}

The $R_{\rm{CP}}$(0-10$\%$/40-60$\%$) of  $\phi$ mesons at mid-rapidity ($|y| < 0.5$) in
Au+Au collisions at $\sqrt{s_{NN}}$ = 7.7 - 39 GeV are presented in
the panel (a) of Fig. 1.
The $R_{\rm{CP}}$(0-05$\%$/40-60$\%$) at $\sqrt{s_{NN}}$ = 200 GeV are taken from 
previous STAR measurements~\cite{phi_star_prc}. The $R_{\rm{CP}}$ is
defined as the ratio of the particles yield in the central to
peripheral collisions normalized by number of binary
collisions ($N_{\rm{bin}}$). The value of  $N_{\rm{bin}}$ is calculated from the
Monte Carlo Glauber simulation~\cite{Nbin}. If $R_{\rm{CP}}$ is equal to one, then the
nucleus nucleus collision is simply superposition of nucleon nucleon
collisions. Deviation of $R_{\rm{CP}}$ from the unity would imply
contribution from the nuclear medium effects. Because of the energy
loss of the partons traversing the high density QCD medium the
$R_{\rm{CP}}$ of $\phi$ mesons goes below unity at 200 GeV~\cite{white}.
 From the Fig. 1 one can
see at the intermediate $p_{T}$, $R_{\rm{CP}}$ goes above unity with
decrease in beam energy. This indicates that at lower beam energy the parton
energy loss effect could be less important.  
\begin{figure}[!ht]
\begin{center}
\includegraphics[scale=0.23]{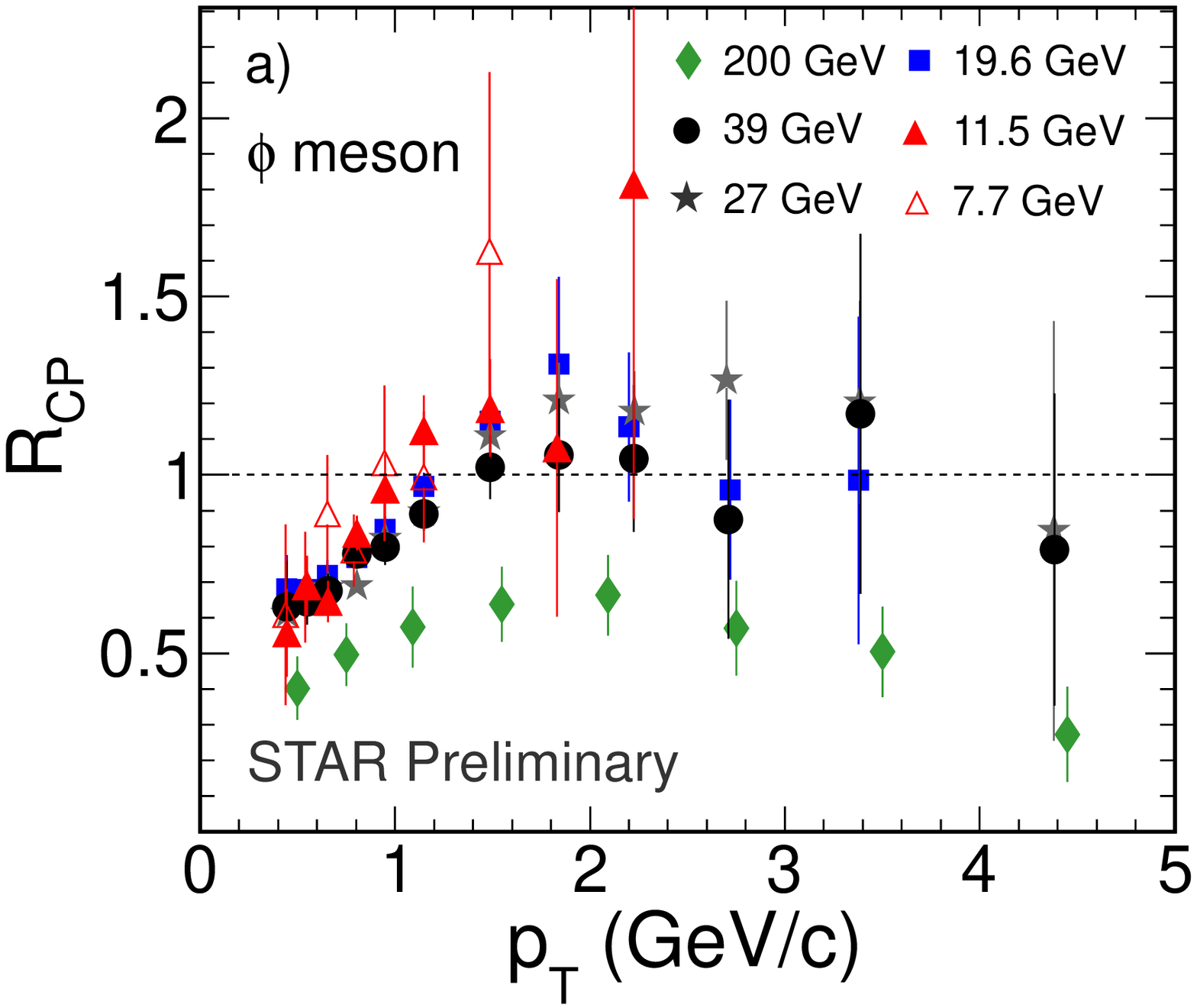}
\includegraphics[scale=0.195]{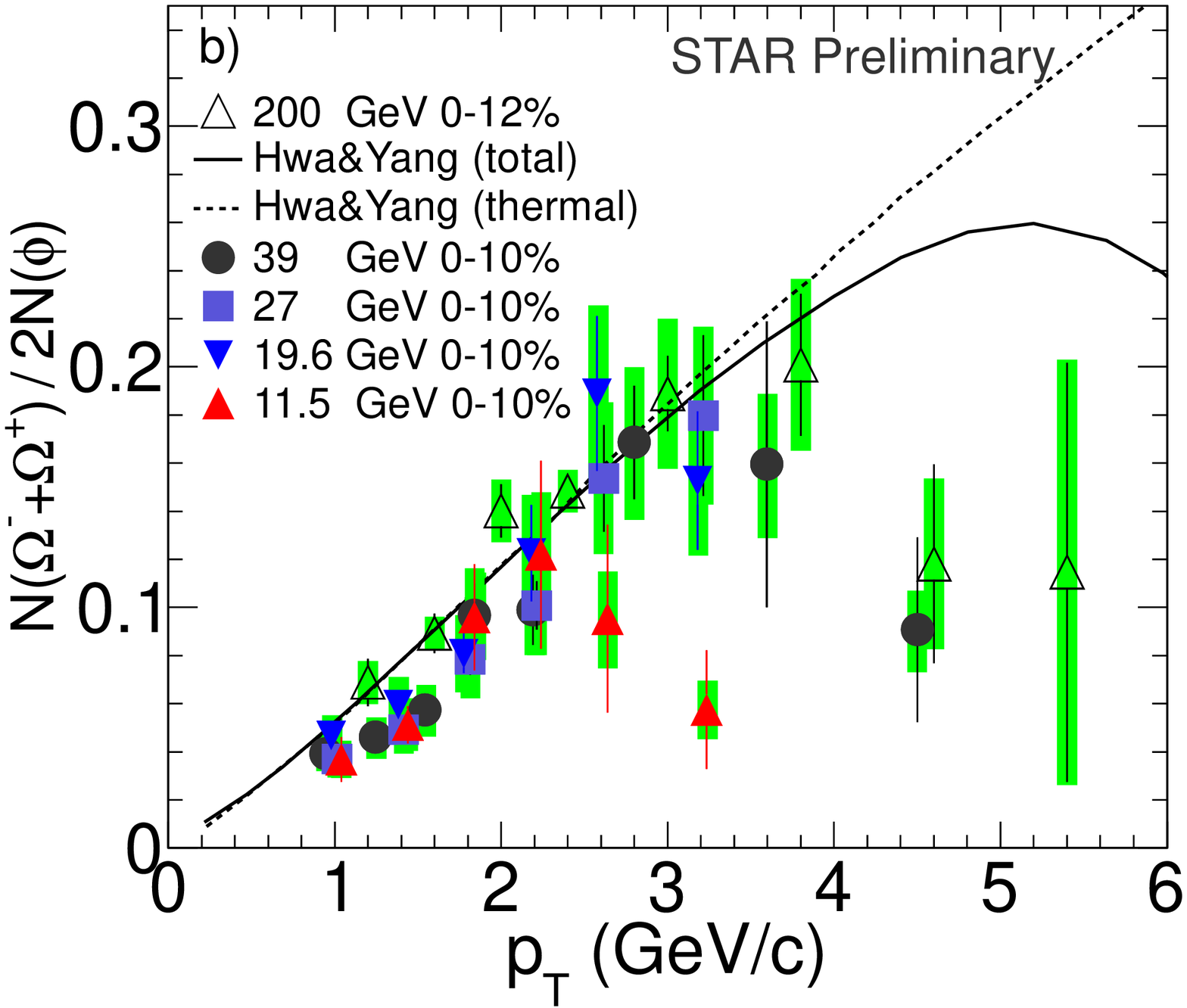}
\caption{(Color online) Panel (a) : The $R_{\rm{CP}}$ as function of
  $p_{T}$ at mid-rapidity ($|y| < 0.5$) in the Au+Au collision
  at various beam energy. Error bars are only
  statistical uncertainties. Panel (b): The baryon-to-meson ratio,
  ($\Omega^{-} + \overline{\Omega}^{+}$)/2$\phi$, as a function of
  $p_{T}$ at mid-rapidity ($|y| < 0.5$). Green
  bands are the systematic errors and vertical lines are statistical errors.}
\label{fig1}
\end{center}
\end{figure} 

The panel (b) of Fig. 1 shows the baryon-to-meson ratio, ($\Omega^{-}
+ \overline{\Omega}^{+}$)/2$\phi$, as a function of $p_{T}$ in Au + Au collisions
at $\sqrt{s_{NN}}$ = 11.5 GeV to 200 GeV. The data points for 200 GeV
are 
taken from Ref.~\cite{phi_star_prc}. The dashed lines are the
results from the recombination model calculations with thermal strange quarks~\cite{thermal_quark}. 
In Au+Au central collisions at $\sqrt{s_{NN}}$ = 200 GeV, the ratios
of ($\Omega^{-} + \overline{\Omega}^{+}$)/2$\phi$ in the intermediate $p_{T}$ range are explained by the
recombination model with thermal strange quarks. The ratios ($\Omega^{-} + \overline{\Omega}^{+}$)/2$\phi$ for 
$\sqrt{s_{NN}}$ $\geq$ 19.6 GeV show similar trend. But at
$\sqrt{s_{NN}}$ = 11.5 GeV, the ratio at the highest measured $p_{T}$
shows a deviation from the trend of other energies.
This may suggest a change in $\Omega$ and/or $\phi$ production
mechanism at $\sqrt{s_{NN}}$ = 11.5 GeV.

It has been observed  from RHIC measurements that when $v_{2}$ and corresponding $p_{T}$ are scaled by number of constituent
quarks of the hadrons, the measured $v_{2}$ values at the intermediate
$p_{T}$ are consistent with expectations from 
parton coalescence or recombination models~\cite{ncq1,ncq1a}. This is known as NCQ scaling and
considered as a signature of partonic collectivity.
\begin{figure}[!ht]
\begin{center}
\centerline{\includegraphics[scale=0.6]{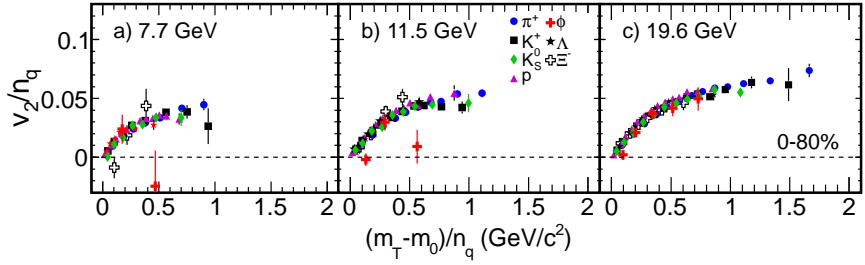}}
\caption{(Color online) The elliptic flow scaled by
  number-of-constituent quark ($n_{q}$) as a
  function of $(m_{T}-m_{0})/n_{q}$ for selected particles in the Au+Au collision
  at  $\sqrt{s_{NN}}$ = 7.7, 11.5 and 19.6 GeV  for 0-80$\%$ centrality~\cite{prc,prl}. Error bars are only
  statistical uncertainties.}
\label{fig5}
\end{center}
\end{figure} 
Figure 2 shows $v_{2}$ divided by number of constituent quark  as
function  of  $(m_{T} - m_{0})/n_{q}$, where $m_{T} =\sqrt{(p_{T}^{2}
+ m_{0}^{2}})$ is the transverse mass and $m_{0}$ is the mass of the hadron, at $\sqrt{s_{NN}}$ = 7.7,
11.5 and 19.6 GeV. The NCQ scaling holds fairly well at $\sqrt{s_{NN}}$ $\geq$  19.6 GeV 
(results for $\sqrt{s_{NN}}$ $>$ 19.6 GeV are not shown here). This
could be considered as a signature of partonic collectivity. However,  we observe
at $\sqrt{s_{NN}}$ = 7.7 and 11.5 GeV  that the $\phi$-meson $v_{2}$
deviates from the
trend of the other hadrons at highest measured  $p_{ T}$
values by 1.8$\sigma$ and 2.3$\sigma$, respectively. Due to the small hadronic interaction cross-section,
 $v_{2}$ of $\phi$ meson mostly reflect collectivity from the partonic
phase~\cite{BN,NBN}. So the small magnitude of the $\phi$-meson $v_{2}$ at
$\sqrt{s_{NN}}$ $\leq$ 11.5 GeV could be the effect for a system, where
hadronic interactions are more important. But more statistics
are needed  at $\sqrt{s_{NN}}$ = 7.7 and 11.5 GeV for $\phi$-meson
$v_{2}$ measurement to draw a clear conclusion and therefore $\phi$
measurement would be one of the focuses in the proposed BES phase II program.

\section{Summary}
We report the study of $\phi$-meson production and elliptic flow at mid-rapidity in Au + Au
collisions at $\sqrt{s_{NN}}$ = 7.7 - 200 GeV recorded by the STAR
detector. At the intermediate $p_{T}$, the nuclear modification factor $R_{\rm{CP}}$ of $\phi$ increases with decreasing beam energies, indicating that the partonic
energy loss effect becomes less important at lower beam
energies. The ratios
of ($\Omega^{-} + \overline{\Omega}^{+}$)/2$\phi$ in the intermediate $p_{T}$ range show a different trend at 11.5 GeV compared to those for the higher beam energies. This may suggest a change of particle
production mechanism at lower beam energy. The NCQ scaling holds for
$\sqrt{s_{NN}}$ $\geq$ 19.6 GeV.
We observe
at $\sqrt{s_{NN}}$ = 7.7 and 11.5 GeV  the $\phi$-meson $v_{2}$ show deviation from the other hadrons at highest measured  $p_{ T}$
values by 1.8$\sigma$ and 2.3$\sigma$, respectively. This may indicate
that the contribution to the collectivity from 
partonic phases decreases at lower beam energies. 

\section{Acknowledgements}
Financial support from DST SwarnaJayanti project, Government of India is gratefully acknowledged. 

\end{document}